\documentclass[pre,aps,twocolumn,showpacs]{revtex4-2}
\usepackage{amsmath,amssymb,graphicx,ulem}
\usepackage{epsfig}
\usepackage{textcomp}
\usepackage[titletoc,title]{appendix}

\usepackage{color}

\begin{document}

\title{Short-time blowup statistics of a Brownian particle in repulsive potentials}

\author{Baruch Meerson}
\email{meerson@mail.huji.ac.il}
\affiliation{Racah Institute of Physics, Hebrew University of Jerusalem, Jerusalem 91904, Israel}

\begin{abstract}
We study the dynamics of an overdamped Brownian particle in a repulsive scale-invariant potential $V(x) \sim -x^{n+1}$. For $n > 1$, a particle starting at position $x$ reaches infinity in a finite, randomly distributed time. We focus on the short-time tail $T \to 0$ of the probability distribution $P(T, x, n)$ of the blowup time $T$ for integer $n > 1$. Krapivsky and Meerson [Phys. Rev. E \textbf{112}, 024128 (2025)] recently evaluated the leading-order asymptotics of this tail, which exhibits an $n$-dependent essential singularity at $T = 0$. Here we provide a more accurate description of the $T \to 0$ tail by calculating, for all $n = 2, 3, \dots$, the previously unknown large pre-exponential factor of the blowup-time probability distribution. To this end, we apply a WKB approximation---at both leading and subleading orders---to the Laplace-transformed backward Fokker--Planck equation governing $P(T, x, n)$. For even $n$, the WKB solution alone suffices. For odd $n$, however, the WKB solution breaks down in a narrow boundary layer around $x = 0$. In this case, it must be supplemented by an ``internal'' solution and a matching procedure between the two solutions in their common region of validity.
\end{abstract}

\maketitle

\section{Introduction}
Finite-time singularities are ubiquitous in nature, and they have attracted much attention from physicists and mathematicians, see e.g. Refs. \cite{MM1975,CP1993,cosmology,EggersRMP,EF2000}. 
When a dynamical quantity reaches infinity in a finite time, we call this singularity a blowup. Probably the simplest model of a blowup is provided by a first-order differential equation
\begin{equation}
\label{DE-n}
\dot{x}(t) = x^n(t)\,,
\end{equation}
where $n>1$. Indeed, for any positive initial condition $x(t=0)\equiv x_0 > 0$, the solution of this equation,
\begin{equation}
\label{determ}
x(t) = \frac{x_0}{\left[1-(n-1) x_0^{n-1} t\right]^{\frac{1}{n-1}}},
\end{equation}
exhibits a blowup at time $t=(n-1)^{-1} x_0^{1-n}$.

Equation~(\ref{DE-n}) describes an overdamped deterministic motion of a particle on the line in a repulsive potential
\begin{equation}\label{potential1d}
U(x,n) = -\frac{x^{n+1}}{n+1}\,.
\end{equation}
In the presence of noise, which we will model as an additive white Gaussian noise with zero mean, Eq.~(\ref{DE-n}) gives way to the Langevin equation
\begin{equation}
\label{SDE-n}
\dot{x}(t) = x^n(t)+ \sqrt{2}\eta(t)\,,n=2,3, \ldots \,,
\end{equation}
with $\langle \eta(t_1)\eta(t_2)\rangle = \delta(t_1-t_2)$. Equation~(\ref{SDE-n}) describes an overdamped motion of a Brownian particle in the power-law repulsive potential \cite{rescaling}.  Here the particle reaches infinity almost surely even for negative $x_0$. For odd $n>1$, the blowups to plus and minus infinity are both possible. For even $n\geq 2$, the blowup is only to $+\infty$. The blowup time is random, and it is interesting to determine its distribution. Mathematically the blowup problem is a particular case of the first passage problem, and one can adapt to this problem some well-developed mathematical tools \cite{Gardiner,SRbook,KRBbook}. 

Different aspects of the model, described by Eq.~(\ref{SDE-n}), have been studied in theory and in experiment \cite{Ornigotti,SilerPRL,Ryabov,KM2025}. 
Besides being a rich and interesting model, it appears to be quite useful. This is because potentials which grow with the distance faster than quadratically successfully model different aspects of a host of stochastic processes in physics and biology.  For example, Eq. \eqref{SDE-n} with  $n=2$ describes the dynamics at the saddle-node bifurcation \cite{Horsthemke,Broeck94} which is encountered in the context of optical bistability in lasers \cite{Colet,laser91}, firing of neurons \cite{neuron-1,neuron-2},  Brownian ratchets \cite{Broeck01,Dean17,Arzola}, and nonlinear maps \cite{Hirsch}. 

As of present, the blowup time probability distribution $\mathcal{P}(T,x_0,n)$ -- or, more precisely, its generating function, or the Laplace transform $T \to s$ -- has been calculated explicitly only for the quartic potential, $n=3$ \cite{KM2025}. Even in that particular case the exact solution -- the inverse Laplace transform of an expression which includes a bi-confluent Heun function, see below -- can only be evaluated numerically. To get a useful insight into this problem it is desirable to obtain analytical asymptotics for the large-$T$ and small-$T$ tails of the distribution $\mathcal{P}(T,x_0,n)$. 

The large-$T$ tail of $\mathcal{P}(T,x_0,n)$ is purely exponential \cite{KM2025}. It is determined by the closest to $s=0$ pole of the generating function or,  in the spectral language, by the ground state of the direct Fokker-Planck operator corresponding to the Langevin equation~(\ref{SDE-n}) \cite{KM2025}. 

The small-$T$ tail of $\mathcal{P}(T,x_0,n)$ is quite different and nontrivial. Here infinitely many terms in the pole decomposition of the generating function (or, equivalently, infinitely many eigenstates of the Fokker-Planck operator) contribute \cite{KM2025}, calling for a different approach. The authors of Ref. \cite{KM2025} evaluated this tail by employing the optimal fluctuation method (OFM) which involves a leading-order saddle-point calculation of the path integral corresponding to the Langevin equation~(\ref{SDE-n}). The OFM is a variant of time-dependent WKB approximation.  The OFM calculation boils down to determining the optimal (that is, most likely) path $x_*(t)$ -- a deterministic trajectory which minimizes the action functional of the path integral conditioned on starting at a specified point $x_0$ and reaching infinity at a specified time $T$ \cite{KM2025}. This calculation yields a leading-order asymptotic of the logarithm of $\mathcal{P}(T\to 0,x_0,n)$. In particular, for $x_0=0$ (a convenient benchmark which we will also adopt in the following) they obtained \cite{KM2025}
\begin{equation}
 \mathcal{P} (T\to 0,n) \sim \exp\left(-\alpha_n\, T^{-\tfrac{n+1}{n-1}}\right)\,, \quad n=2,3,\dots
   \label{smallTleading}
\end{equation}
where 
\begin{equation}
\label{alphan}
  \alpha_n = \frac{\Gamma
   \left(\frac{3}{2}-\frac{1}{2
   n}\right) \Gamma\!\left(\frac{1}{2
   n}\right) \left[\Gamma
   \left(1+\frac{1}{2 n}\right) \Gamma\!\left(\frac{n-1}{2
   n}\right)\right]^{\frac{n+1}{n-1}} }{4 (n+1)\pi^{\frac{n}{n-1}} }>0\,,
\end{equation}
and $\Gamma(\dots)$ is the gamma function. The $n$-dependent essential singularity of the $T\to 0$ tail  is clearly seen from Eq.~(\ref{smallTleading}).  

Equation~(\ref{smallTleading}), however, misses a pre-exponential factor which can be important because of its possible dependence on the small parameter $T$. As we will see shortly, the prefactor does depend on $T$ in this problem, and it is large. The determination of this prefactor demands going beyond the leading-order OFM/WKB calculation. 

In the remainder of this paper  we provide an accurate description of the $T \to 0$ tail of the blowup time distribution $\mathcal{P}(T,x_0,n)$
by explicitly calculating, for all $n=2,3,\dots$, this previously unknown 
pre-exponential factor. To achieve this goal we employ a different version of the WKB method, in the leading and subleading orders, by applying it to the Laplace-transformed backward Fokker--Planck equation, which describes the generating function 
of $\mathcal{P}(T,x_0,n)$. This variant of WKB theory is advantageous because -- by virtue of the Laplace transform and in contrast to the time-dependent OFM --  it deals with a time-independent problem.

As we will see, for even $n$ the (leading and subleading) WKB solution provides a valid asymptotic for all $x$, and it suffices for an accurate calculation of the prefactor. For odd $n$, however,
the WKB solution becomes invalid in a small boundary layer around $x=0$. Here we proceed by finding a separate, ``internal" solution,  which then can be matched to the WKB 
solution in their common region of validity. 

Here is a plan of the remainder of the paper. In Sec. \ref{generating} we recap the exact formulation of the problem. The WKB solution, the internal solution and their matching are presented in Sec. \ref{theory}. Finally, Sec. \ref{discussion} includes a brief summary and discussion of our main results.

\section{Generating function  of $\mathcal{P}(T)$} 
\label{generating}
   
We are interested in the small-$T$ asymptotic of the blowup time distribution $\mathcal{P}(T,x,n)$, where $n>1$ is integer, and $x$ is the initial position of the particle (we have dropped the subscript $0$ of $x_0$).  The generating function of this distribution coincides with its Laplace transform:
\begin{equation}
\label{LT}
\Pi(s,x,n)=\left\langle e^{-s T(x)}\right\rangle = \int_0^\infty dT\,e^{-sT} \mathcal{P}(T,x,n)\,.
\end{equation}
It is known that this function obeys the ordinary differential equation (ODE)
\begin{equation}
\label{Pi-eq}
\Pi^{\prime\prime}(x) +x^n\,\Pi^{\prime}(x) = s \Pi(x)\,,
\end{equation}
obtained by applying the Laplace transform $T\to s$ to the backward Fokker-Planck equation for the Markovian stochastic process described by the Langevin equation~(\ref{SDE-n}), see \textit{e.g.} Refs.~\cite{Gardiner,SRbook,KRBbook,KR18}. The first passage to $x=+\infty$ corresponds to the boundary 
conditions
\begin{equation}
\label{BCslaplace}
\Pi(x\to+\infty)=1 \quad \text{and} \quad \Pi^{\prime}(x \to-\infty)=0\,.
\end{equation} 
The primes  in Eqs.~(\ref{Pi-eq}) and (\ref{BCslaplace}) denote the derivatives with respect to $x$, and we have dropped some of the arguments of $\Pi(s,x,n)$. 

When $n$ is odd, the Laplace transform $\Pi(s,x,n)$ possesses the symmetry $\Pi(s,-x,n)=\Pi(s,x,n)$, which implies
\begin{equation}
\label{BCsymmetric}
\Pi^{\prime}(x=0)=0\,.
\end{equation} 
Therefore, for odd $n$ we can limit ourselves to  $x>0$ and  
effectively replace the second boundary condition in Eq.~(\ref{BCslaplace}) by the symmetry condition \cite{footnoteodd}
\begin{equation}
\label{BCsymmetric}
\Pi^{\prime}(x=0)=0\,.
\end{equation} 
As we will see shortly, this fact has important consequences.

Once the generating function $\Pi(s,x,n)$ is determined,  the blowup time probability distribution $\mathcal{P}(T,x,n)$ is given  by the inverse Laplace transform of $\Pi(s,x,n)$:
\begin{equation}
\label{inversetr}
 \mathcal{P}(T,x,n) = \frac{1}{2\pi i}\int_{\gamma-i \infty}^{\gamma+i \infty} e^{sT} \Pi(x,s,n) \,ds\,.
\end{equation}

As of present, an exact analytical solution of the problem described by Eqs.~(\ref{Pi-eq}) and  (\ref{BCslaplace}) has been obtained only for a quartic potential, $n=3$. This solution has the form \cite{KM2025}
\begin{equation}
\label{Pi-sol}
\Pi(s,x)=\frac{\text{HeunB}(s/4, 0, 1/2, 0, 1/2; x^2)}{\text{HeunB}(s/4, 0, 1/2, 0, 1/2; \infty)}\,,\quad n=3,
\end{equation}
where $\text{HeunB}(a_1,a_2,a_3,a_4,a_5; z)$ is the bi-confluent Heun function \cite{NIST}.
In the benchmark case $x=0$ (the particle starts at $x=0$) the generating function \eqref{Pi-sol} simplifies to
\begin{equation}
\label{Pi-0}
\Pi_0(s)=\frac{1}{\text{HeunB}(s/4, 0, 1/2, 0, 1/2; \infty)}\,,\quad n=3.
\end{equation}

The $T\to 0$ asymptotic of $\mathcal{P}(T,x)$ is determined by the $s\to \infty$ asymptotic of the  bi-confluent Heun function,  which does not seem to be  available in the literature. Here we will determine the $s\to \infty$ asymptotic of $\Pi(s,x)$ directly from Eqs.~(\ref{Pi-eq})-(\ref{BCsymmetric}), and we will do it for all $n=2,3,\dots$. To this end we will develop a suitable WKB approximation in the leading and subleading orders and, when necessary,  supplement it  with an asymptotic ``internal" solution and a matching procedure. The whole method relies on the large parameter $s\gg 1$ which corresponds to $T\ll 1$.

\section{WKB theory, internal region and matching}
\label{theory}

\subsection{WKB solution}
The WKB approximation, see e.g. Ref. \cite{BenderOrszag}, involves the exponential ansatz
\begin{equation}\label{WKBansatz}
\Pi(x) = \exp\left[-S_0(x)-S_1(x)\right]\,.
\end{equation}
Plugging it into Eq.~(\ref{Pi-eq}), we obtain an exact equation,
\begin{equation}\label{exacteq}
 \left(S_0^{\prime}+S_1^{\prime}\right)^2-\left(S_0^{\prime\prime}+S_1^{\prime\prime}\right) -x^n \left(S_0^{\prime}+S_1^{\prime}\right)=s\,.
\end{equation} 
Now we assume that the parameter $s\gg 1$ and search for a solution such that $S_1(x) \ll S_0(x)$. Equation~(\ref{exacteq}) includes three types of terms: the leading-order terms, the subleading-order terms, and the terms that can be neglected even in the subleading order. In the leading order we can neglect $S_0^{\prime\prime}$ and all the terms which involve $S_1$. This leads to the equation
\begin{equation}\label{leadeq}
\left(S_0^{\prime}\right)^2-x^n S_0^{\prime}-s = 0\,,
\end{equation}
which has the form of a Hamilton-Jacobi equation for the time-independent Hamiltonian 
\begin{equation}\label{H0}
  H(x,p) = p^2-x^n p\,.
\end{equation}
Here $x$ and $p$ are the canonical variables, and $s$ can be interpreted as the (conserved) energy. Not surprisingly, the Hamilton-Jacobi description of this problem in the leading order of the WKB theory is equivalent to the Lagrangian description  of the same problem in the framework of the OFM used in  Ref. \cite{KM2025}.  

The first-order ODE (\ref{leadeq}) is a quadratic equation for $S_0^{\prime}$. To obey the first boundary condition in Eq.~(\ref{BCslaplace}), we must choose the solution with the minus sign:
\begin{equation}\label{S0prime}
S_0^{\prime}(s,x) = \frac{x^n-\sqrt{x^{2n}+4s}}{2}\,.
\end{equation}
At this stage we notice an important difference between the cases of even and odd $n$.  For odd $n$ the symmetry condition~(\ref{BCsymmetric}) translates into the two following conditions
\begin{equation}\label{Sat0}
S_0^{\prime} (x=0) = S_1^{\prime}(x=0) = 0\,.
\end{equation}
Equation~(\ref{S0prime}) clearly violates the first of these two conditions. This signals that, for odd $n$,
the WKB approximation is invalid in a small region around $x=0$, where a different (non-WKB) solution should take care of the boundary condition~(\ref{S0prime}).  In the next subsection we will indeed find such a solution -- the internal solution -- for odd $n$, and match it to the WKB solution in their common region of validity. 

In the subleading WKB order  we obtain the following equation:
\begin{equation}\label{S1prime}
S_1^{\prime}(x) = \frac{p^{\prime}(x)}{2p(x)-x^n}\,,
\end{equation}
where we have denoted for brevity $S_0^{\prime} (s) \equiv p(x)$. To obtain Eq.~(\ref{S1prime}) from Eq.~(\ref{exacteq}) we used Eq.~(\ref{leadeq}) and neglected the terms $(S_1^{\prime})^2$ and $S_1^{\prime\prime}$, which are of a sub-subleading order. 

Integrating Eq.~(\ref{S0prime}) over $x$ and using the condition $S_0(s,x\to \infty)=0$, which follows from the first  boundary condition in Eq.~(\ref{BCslaplace}), we obtain
\begin{eqnarray}
\label{S0sol}
S_0(x) &=&-\sqrt{s} \,x \, _2F_1\left(-\frac{1}{2},\frac{1}{2 n};1+\frac{1}{2 n};-\frac{x^{2 n}}{4
   s}\right)+\frac{x^{n+1}}{2 n+2}\nonumber\\
   &-&\frac{2^{\frac{1}{n}-1} s^{\frac{1}{2 n}+\frac{1}{2}} \Gamma \left(1+\frac{1}{2
   n}\right) \Gamma \left(-\frac{n+1}{2 n}\right)}{\sqrt{\pi }}\,,
\end{eqnarray}
%\label{S0sol}
%  S_0(x) &=& \frac{n x^{n+1} \, _2F_1\left(\frac{1}{2},-\frac{n+1}{2 n};\frac{n-1}{2 n};-\frac{4 s}{x^{2 n}}\right)}{2
%   (n+1)}\nonumber\\
%   &-&\frac{1}{2} x \sqrt{x^{2 n}+4 s}+\frac{x^{n+1}}{2 (n+1)}\,,
%\end{eqnarray}
where $\,_2F_1(\dots)$ is the hypergeometric function. Using this leading-order solution, we can rewrite the subleading-order equation~(\ref{S1prime}) in an explicit form:
\begin{equation}\label{S1primemore}
S_1^{\prime}(x) = \frac{n x^{n-1} \left(x^n-\sqrt{x^{2 n}+4 s}\right)}{2 \left(x^{2 n}+4 s\right)}\,.
\end{equation}
Integrating this equation over $x$ and demanding that $S_1(s,x\to \infty)=0$, we obtain
\begin{equation}\label{S1sol}
S_1(x) \!=\!\frac{1}{2} \ln \left( x^{2 n}+4 s-x^n\sqrt{x^{2 n}+4 s}\right)-\ln \sqrt{2 s}.
\end{equation}
Overall, the approximate WKB solution for $\Pi(s,x)$ is given by Eq.~(\ref{WKBansatz}) with
\begin{eqnarray}\label{S} 
 S(x) &\equiv& S_0(x)+S_1(x) \nonumber \\
 &=&-\sqrt{s} \,x \, _2F_1\left(-\frac{1}{2},\frac{1}{2 n};1+\frac{1}{2 n};-\frac{x^{2 n}}{4s}\right) \nonumber\\
 &-&\frac{2^{\frac{1}{n}-1} s^{\frac{1}{2 n}+\frac{1}{2}} \Gamma 
 \left(1+\frac{1}{2n}\right) \Gamma \left(-\frac{n+1}{2 n}\right)}{\sqrt{\pi }} +\frac{x^{n+1}}{2 (n+1)}\nonumber\\
 &+&\frac{1}{2} \ln \left( x^{2 n}+4 s-x^n\sqrt{x^{2 n}+4 s}\right) -\ln \sqrt{2 s}\,.
\end{eqnarray}
%\begin{eqnarray}\label{S}
% S(x) &\equiv& S_0(x)+S_1(x) \nonumber \\
% &=&\frac{n x^{n+1} \, _2F_1\left(\frac{1}{2},-\frac{n+1}{2 n};\frac{n-1}{2 n};-\frac{4 s} {x^{2 n}}\right)}{2(n+1)}\nonumber \\
% &+& \frac{x^{n+1}}{2 (n+1)}-\frac{1}{2} x \sqrt{x^{2 n}+4 s}-\ln \sqrt{2 s}
%  \nonumber \\ 
%   &+&\frac{1}{2} \ln \left(x^{2 n}+4
%   s-x^n \sqrt{x^{2 n}+4 s}\right)\,.
%\end{eqnarray}

\subsection{Even $n$: the WKB solution holds for all $x$}

\begin{figure}
\includegraphics[width=0.35\textwidth,clip=]{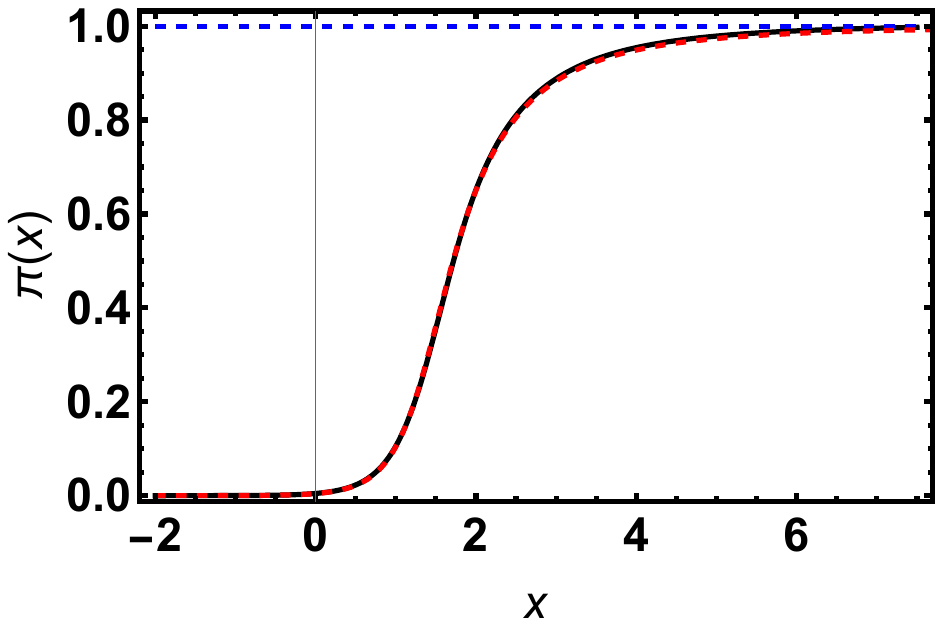}
\caption{Solid line: numerical solution of Eq.~(\ref{Pi-eq}) for the generating function for $n=4$ and $s=10$ \cite{artrelaxation}. Red dashed line:
the WKB asymptotic given by Eqs.~(\ref{WKBansatz}) and~(\ref{S}). Blue dashed line: the asymptotic $\Pi(x\to \infty)=1$.}
\label{numsol}
\end{figure}

By inspecting Eqs.~(\ref{S0sol}) and (\ref{S1sol}) for $S_0$ and $S_1$, one can see that, for sufficiently large $s$, the assumptions we made about the leading and subleading WKB orders are valid for all $x$ once $n$ is even. 
In this case Eqs.~(\ref{WKBansatz}) and~(\ref{S}) readily yield the desired large-$s$ asymptotic of the generating function $\Pi(x)$. 
As an illustration, Fig.~\ref{numsol} compares this WKB asymptotic with a numerical solution of the exact equation~(\ref{Pi-eq}) with the boundary conditions~(\ref{BCslaplace}) for $n=4$ \cite{artrelaxation}. As one can see from Fig. \ref{numsol}, the agreement is quite good already for $s=10$.

%\vspace{0.5cm}

In our benchmark case $x=0$ the WKB solution~(\ref{WKBansatz}) and~(\ref{S}) yields
\begin{equation}\label{PIeven}
  \Pi_0(s\gg 1,n) \simeq \frac{1}{\sqrt{2}}\,e^{-A_n s^{\frac{n+1}{2 n}}}\,,\; n=2k, \;k=1,2,\dots\,,
\end{equation}
where
\begin{equation}\label{An}
A_n=-\frac{\Gamma\left(1+\frac{1}{2 n}\right) \Gamma\left(-\frac{n+1}{2 n}\right)}{\sqrt{\pi} \,2^{\frac{n-1}{n}} }>0\,.
\end{equation}

\subsection{Odd $n$: internal region and matching}

\begin{figure}
\includegraphics[width=0.35\textwidth,clip=]{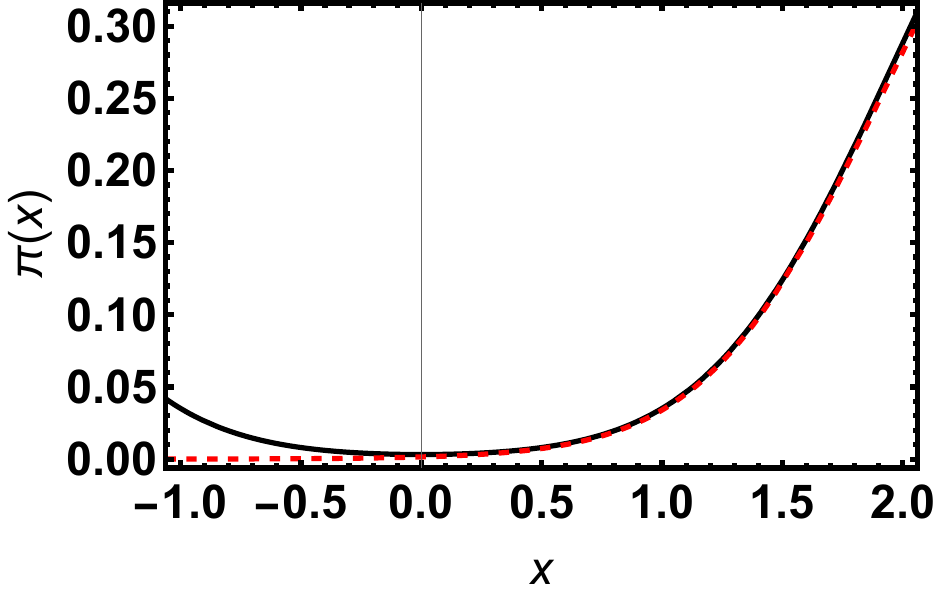}
\caption{Solid line: the exact generating function $\Pi(x)$ as described by Eq.~(\ref{Pi-sol}) for $n=3$ and $s=10$. Red dashed line: the WKB asymptotic of $\Pi(x)$ given by Eqs.~(\ref{WKBansatz}) and~(\ref{S}).}
\label{comparison3}
\end{figure}

For $n=3,5,7,\dots$ the WKB approximation is invalid in a small boundary layer, $x\lesssim s^{-1/2}$. This is illustrated by Fig.~\ref{comparison3}, which compares the WKB asymptotic (\ref{WKBansatz}) and~(\ref{S}) for $n=3$ and $s=10$ with the exact generating function, given by Eq.~(\ref{Pi-sol}). Here we focus on relatively small $x$. As one can see, the WKB asymptotic starts to deviate from the exact one at small positive $x$. (By construction, this WKB solution does not apply for negative $x$ \cite{negativex}.)

Therefore, we proceed to constructing an approximate analytic solution that would describe the boundary layer  $x\lesssim s^{-1/2}$. This boundary layer  is located inside a much larger ``internal" region, which is defined by the condition $|x|\ll s^{\frac{1}{2n}}$.

As one can check a posteriori, in the internal region the second term on the left in Eq.~(\ref{Pi-eq}) is much smaller than the other two terms of the equation. Neglecting this term, we arrive at  the approximate equation 
\begin{equation}\label{BLeq}
\Pi^{\prime\prime}(x)=s \Pi(x)\,,
\end{equation}
which is independent of $n$ and very simple. Its solution, satisfying the boundary condition~(\ref{BCsymmetric}), is 
\begin{equation}\label{BLlarge}
 \Pi_{<}(x) = C\,\cosh (\sqrt{s}x)\,,
\end{equation}
where $C=C(s,n)$ is a yet unknown function of $s$ and $n$.  The applicability domain of the internal solution, $|x|\ll s^{\frac{1}{2n}}$, is determined by the demand that the neglected term $x^n \Pi^{\prime}(x)$  is indeed small. 

To determine the function $C(s,n)$, we consider the $x \gg 1/\sqrt{s}$ asymptotic of the internal solution~(\ref{BLlarge}):
\begin{equation}\label{BLasymp}
  \Pi_{<}(x\gg 1/\sqrt{s}) \simeq \frac{C}{2} \,e^{\sqrt{s}x}\,.
\end{equation}
In its turn, the $x\ll s^{\frac{1}{2n}}$ asymptotic of the WKB solution  (\ref{S}) is the following:
\begin{equation}\label{Ssmallx}
S\left(x\ll s^{\frac{1}{2n}}\right) \simeq  -\sqrt{s} x +A_n s^{\frac{n+1}{2 n}}+\ln \sqrt{2}\,,
\end{equation}
where $A_n$ has been defined in Eq.~(\ref{An}). The resulting asymptotic of $\Pi(x)$ 
is 
\begin{eqnarray}\label{Pismallx}
\Pi_{\text{WKB}}\left(x\ll s^{\frac{1}{2n}}\right) &\simeq& \exp\left[-S\left(x\ll s^{\frac{1}{2n}}\right)\right] \nonumber \\ 
&\simeq&
\frac{1}{\sqrt{2}}e^{\sqrt{s} x -A_n s^{\frac{n+1}{2 n}}}\,.
\end{eqnarray}
As we can see,  the asymptotics~(\ref{BLasymp}) and~(\ref{Pismallx}) can be matched
in their common region of validity $s^{-\frac{1}{2}}\ll x \ll s^{\frac{1}{2n}}$, and we obtain 
\begin{equation}\label{C(s)}
C(s,n) = \sqrt{2} e^{-A_n s^{\frac{n+1}{2 n}}}\,,
\end{equation}
which completes our solution for $\Pi(x)$ for odd $n$ and $s\gg 1$. 

Applying Eqs.~(\ref{BLlarge}) and (\ref{C(s)}) to our benchmark case $x=0$ (notice that $x=0$ is located inside the boundary layer and outside the WKB region!), we obtain
\begin{equation}\label{Pi_0}
\Pi_0(s\gg 1,n) = C(s,n) =  \sqrt{2} e^{-A_n s^{\frac{n+1}{2 n}}}
\end{equation}
for $n=2k+1$, $k=1,2,\dots$. As one can see, Eq.~(\ref{Pi_0}) almost coincides with
the expression in Eq.~(\ref{PIeven}) obtained for even $n$. The only difference from Eq.~(\ref{PIeven}) is 
an additional factor $2$ which we discuss below. 

%\vspace{0.5cm}
\begin{figure}
\includegraphics[width=0.35\textwidth,clip=]{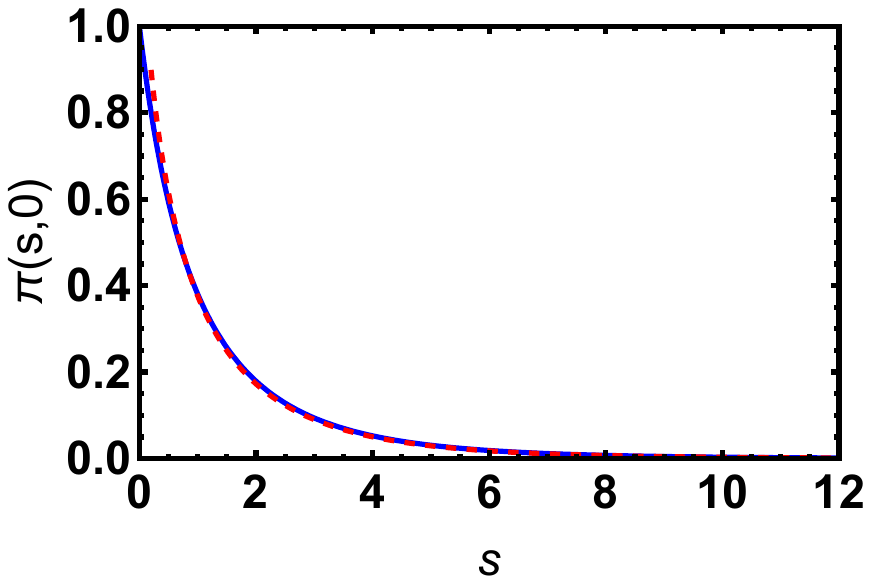}
\caption{Solid line: the exact expression~(\ref{Pi-0}) for the generating function $\Pi(s,0)$ for $n=3$. Dashed line:
the large-$s$ asymptotic of $\Pi(s,0)$, given by Eq.~(\ref{const}).}
\label{Pi(s)}
\end{figure}

In particular, for the quartic potential $n=3$ Eq.~(\ref{Pi_0}) yields
\begin{equation}\label{const}
\Pi_0(s\gg 1,n=3)=\sqrt{2}\, e^{-\frac{\Gamma(1/6) \Gamma(1/3) s^{2/3} }{2^{8/3}\sqrt{\pi }}}\,.
\end{equation}
Figure \ref{Pi(s)} compares this $s\gg 1$ asymptotic with the exact expression (\ref{Pi-0}) for $\Pi_0(s)$. As one can see, the agreement is quite good even for moderate $s$.

\vspace{0.2cm}
\subsection{Short-time tail of $\mathcal{P}(T,n)$}

The $T \to 0$ tail of the blowup-time distribution $\mathcal{P}(T,x=0,n)$ is determined by the inverse Laplace transform (\ref{inversetr}), with the generating function $\Pi_0(x=0,s,n)$ approximated by its $s\gg 1$ asymptotic~(\ref{PIeven}) for even $n$ or~(\ref{Pi_0}) for odd $n$.  The large parameter $s$ calls for the steepest-descent evaluation of the integral in Eq.~(\ref{inversetr}). A standard steepest-descent calculation yields, for any integer $n>1$ \cite{footnoteinvLaplace},
\begin{equation}\label{PsmallT}
  \mathcal{P}(T\to 0,n)\simeq
\mu_n \kappa_n\; T^{-\tfrac{3n-1}{2(n-1)}}\!
\exp\!\left(-\beta_n\, T^{-\tfrac{n+1}{n-1}}\right).
\end{equation}
Here  $\mu_n=1$ for even $n$, and $2$ for odd $n$, and the $n$-dependent constants $\kappa_n$ and $\beta_n$ are the following:
\begin{eqnarray}
  \kappa_n &=& \frac{\left[\Gamma \left(\frac{2n+1}{2 n}\right)\right]^{\frac{n}{n-1}}\left[\Gamma \left(\frac{n-1}{2 n}\right)\right]^{\frac{3n-1}{2(n-1)}}}
  {4 \pi ^{\frac{2n-1}{2(n-1)}}  \left[\Gamma\left(\frac{3n-1}{2n}\right)\right]^{1/2}}, \label{D}\\
  \beta_n &=& \frac{(n-1)  \left[2^{\frac{1}{n}-1} \Gamma \left(1+\frac{1}{2n}\right) \Gamma \left(\frac{n-1}{2 n}\right)\right]^{\frac{2 n}{n-1}}}{\pi ^{\frac{n}{n-1}}(n+1)}.\label{B}
\end{eqnarray}

The presence in Eq.~(\ref{PsmallT}) of the factor $\mu_n=2$ for odd $n$ is explained by the fact that, in this case and for $x=0$, a blowup is equally probable to both $+\infty$ and $-\infty$ \cite{KM2025}.  It is quite interesting that the $T\to 0$ asymptotic solution accommodates this feature by producing a small boundary layer around $x=0$ which is captured by the internal solution, and where the WKB approximation breaks down.

Equation~(\ref{PsmallT}) exactly reproduces the character of essential singularity at $T=0$ previously determined in Ref. \cite{KM2025}. Furthermore, 
using gamma functions identities, one can verify that the coefficient $\beta_n$ in Eq.~(\ref{B}) perfectly coincides, for all $n$, with the coefficient $\alpha_n$ in Eq.~(\ref{alphan}), previously calculated in Ref. \cite{KM2025}.  The new asymptotic~(\ref{PsmallT})-(\ref{B}), however, is considerably more accurate than the one described by Eqs.~(\ref{smallTleading}) and (\ref{alphan}), 
because it also accounts for the large pre-exponential factor $\sim T^{-\tfrac{3n-1}{2(n-1)}}$ and an $n$-dependent numerical coefficient $O(1)$.

Figure~\ref{P(T,3} compares, for $n=3$, the short-time asymptotic (\ref{PsmallT}) with the exact distribution obtained by a numerical evaluation of the inverse Laplace transform of  $\Pi_0(s)$ from Eq.~(\ref{Pi-0}). (To remind the reader, $n=3$ is the only case where an explicit and exact analytical solution is available \cite{KM2025}.) As one can see, the agreement is very good. For completeness, Fig.~\ref{P(T,3}   also shows the long-time asymptotic $\mathcal{P}(T) \simeq 2.5\, e^{-1.3685\dots T}$ calculated in Ref. \cite{KM2025}.

\begin{figure}
\includegraphics[width=0.35\textwidth,clip=]{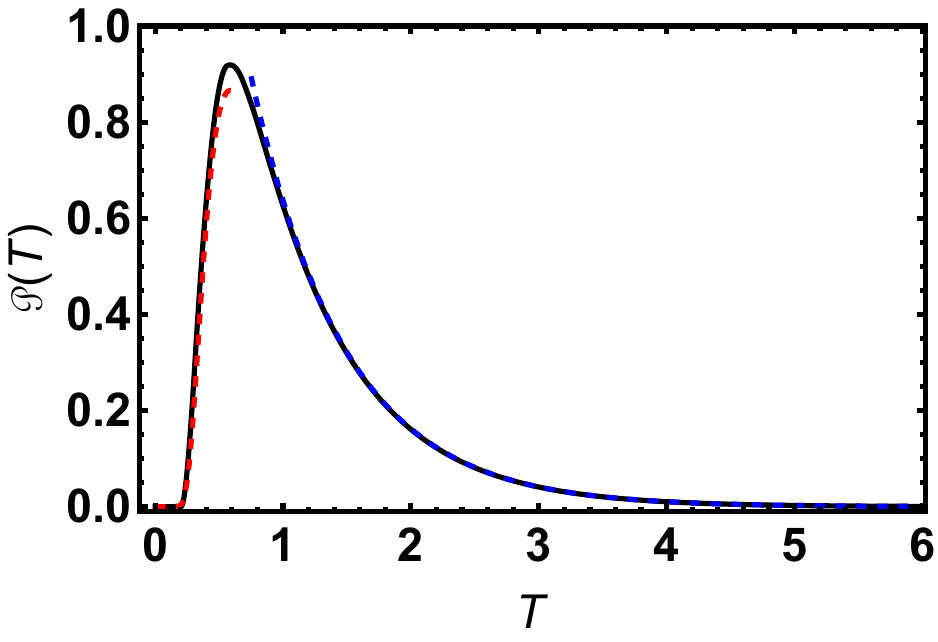}
\caption{Solid line: the blowup time probability distribution $\mathcal{P}(T)$ for $n=3$ and $x=0$, obtained by a numerical inverse Laplace transform of the exact generating function~(\ref{Pi-0}). The red dashed line shows the short-time asymptotic (\ref{PsmallT}). The blue dashed line shows the long-time asymptotic  $\mathcal{P}(T) \simeq 2.5 \, e^{-1.3685\dots T}$ obtained in Ref. \cite{KM2025}.}
\label{P(T,3}
\end{figure}

\section{Summary and Discussion}
\label{discussion}

We developed an accurate description of the $T \to 0$ tail of the blowup time distribution 
by determining, for all $n=2,3,\dots$, the previously unknown $T$-dependent
pre-exponential factor which turns out to be large and therefore especially important. We employed a WKB method, in the leading and subleading orders, to solve the Laplace-transformed backward Fokker--Planck equation, which describes the generating function 
of $\mathcal{P}(T,x_0,n)$. The solution is greatly facilitated by the fact that, in contrast to the time-dependent variants of the WKB theory,  here one deals with a time-independent problem. 

We have shown that for even $n$ the (leading and subleading) WKB approximation alone suffices for the accurate calculation of the prefactor. For odd $n$ 
the WKB solution must be matched to a different solution in a large ``internal" region which includes a small boundary layer around $x=0$  where the WKB approximation breaks down. The matching procedure is possible here because of the presence of a common region of validity where the two approximations overlap. 

In conclusion, the asymptotic method for determining the short-time tail of $\mathcal{P}(T)$, that we proposed here, is quite general and extendable to a whole class of first-passage problems in Markovian systems.

\bigskip
\noindent
{\bf Acknowledgment}. This research was supported by the Israel Science Foundation (Grant No. 1499/20).


\begin{thebibliography} {99}
\bibitem{rescaling} The dimensionless form of Eqs.~(\ref{DE-n})-(\ref{SDE-n}) is obtained by rescaling time and the coordinate.
\bibitem{artrelaxation} Employing the method of artificial relaxation, we introduced an artificial time variable, 
$\tau$, and replaced Eq.~(\ref{Pi-eq}) with the partial differential equation (PDE) 
$$
\;\;\;\partial_{\tau} \Pi(x,\tau) = \partial_{x}^2 \Pi(x,\tau) +x^n \partial_x\Pi(x,\tau) - s \Pi(x,\tau) \,.
$$
The steady-state solution of this equation is described by Eq.~(\ref{Pi-eq}). The PDE was solved numerically with the boundary conditions $\Pi(x=-L_1,\tau) = 0$ and $\Pi(x=L_2,\tau) = 1$  and a step initial condition $\Pi(x,\tau=0) = \theta(x)$ on a sufficiently large finite interval $-L_1<x<L_2$. The solution converges to the steady state, which we are after, quite quickly.%on a time scale $\tau=O(1)$.
\bibitem{negativex} For $x<0$ correct WKB solution can be obtained by choosing the plus sign in front of the square root in Eq.~(\ref{S0prime}). This is equivalent to a mirror reflection of the WKB solution for  $x>0$  with respect to the origin: $\Pi(-x)=\Pi(x)$.  Still, for odd $n$, the WKB asymptotic is valid only outside the boundary layer $|x|\lesssim s^{-1/2}$.
\bibitem{footnoteinvLaplace} We used the principal branch of the fractional power $s^{\frac{n+1}{2n}}$ in the complex plane, so that the branch cut lies along the negative real axis, while the Bromwich contour and its steepest–descent deformation remain entirely  in the analytic half-plane $\mathcal{R}(s)>0$.
\end{thebibliography}
\end{document}